\documentclass[aps,nofootinbib,prd,eqsecnum,showpacs,showkeys,preprintnumbers]{revtex4-2}
\usepackage[caption=false]{subfig}
\usepackage{graphicx}
\usepackage{amsmath}
\usepackage{amsfonts}
\usepackage{amssymb}
\usepackage{float}
\usepackage{color}
\usepackage{bm}
\usepackage{mathrsfs}
\usepackage{epstopdf}
\usepackage{url}
\usepackage{footnote}
\usepackage{textcomp}
\usepackage{ulem}
\usepackage{esint}
\usepackage[unicode=true, pdfusetitle,
 bookmarks=true,bookmarksnumbered=false,bookmarksopen=false,
 breaklinks=false,pdfborder={0 0 1},backref=false,colorlinks=false]{hyperref}
\usepackage{multirow}
\usepackage{pifont}

\begin{document}

\title{Leptogenesis Effects on the Gravitational Waves Background:
Interpreting the NANOGrav Measurements and JWST Constraints on Primordial
Black Holes}
\author{K. El Bourakadi$^{1}$}
\email{k.elbourakadi@yahoo.com}
\author{H. Chakir$^{1}$}
\email{chakir10@gmail.com }
\author{M. Yu. Khlopov$^{2,3,4}$}
\email{khlopov@apc.in2p3.fr}
\date{\today }

\begin{abstract}
We demonstrate that the leptogenesis mechanisms, which are associated with
B-L symmetry breaking mechanism has notable effects on the production of
gravitational waves. These gravitational waves align well with the recent
observations of a stochastic gravitational wave background by NANOGrav and
pulsar-timing arrays (PTAs). For these gravitational waves to match the
recent measurements, the critical value of the\ B-L breaking should be
around the GUT scale. Moreover, we consider the generation of primordial
gravitational waves from binary systems of Primordial\ Black Holes (PBHs)
which could be predicted by the recent detection of gravitational waves.
PBHs with specific masses can be responsible for massive galaxy formation
observed at high redshifts reported by the James Webb Space Telescope
(JWST). We contemplate the potential for a shared source between the
NANOGrav and JWST observations, namely primordial black holes. These black
holes could serve as seeds of rapid galaxy formation, offering an
explanation for the galaxies observed by JWST.
\end{abstract}


\affiliation{$^{1}${\small Subatomic Research and Applications Team, Faculty of Science
Ben M'sik,}\\
{\small Casablanca Hassan II University, Morocco}\\} 
\affiliation{$^{2}$
Institute of Physics, Southern Federal University, 194 Stachki, Rostov-on-Donu, Russia\\}
\affiliation{$^{3}$
Virtual Institute of Astroparticle Physics, 75018 Paris, France\\} 
\affiliation{$^{4}$ 
National Research Nuclear University MEPhI, 115409 Moscow, Russia\\}

\maketitle

\section{Introduction}


Gravitational Waves Background (GWB) provides a captivating glimpse into the
early stages of the universe \cite{I1}. These waves emerge either from
quantum fluctuations occurring during the inflationary period \cite{I2},
sourced by the phase transition of the Universe, results from cosmic strings 
\cite{I3,I4}, or Binary Black Holes (BBH) from the early Universe \cite{I5}.

During the B-L breaking phase transition ending hybrid inflation, most of
the vacuum energy density gets swiftly transferred to B-L Higgs bosons in a
non-relativistic state and a significant portion also goes into the
formation of cosmic strings. Entropy and baryon asymmetry are produced
through thermal and nonthermal leptogenesis as a result of the decay
processes involving massive Higgs bosons and Majorana neutrinos \cite{I6,I7}.

The operators responsible for violating the lepton (or B-L) number, emerge
as a result of the exchange of massive Majorana neutrinos $N_{i}$ (where $%
i=1,2,3$) \cite{U1,U2}. When these heavy neutrinos $N_{i}$ undergo decay,
they inherently generate lepton (or B-L) asymmetry, assuming that the
principles of C and CP conservation do not hold [8]. The resulting lepton
asymmetry is then converted into baryon asymmetry during the early stages of
the universe \cite{U3}. As a consequence, the process of leptogenesis
appears to be the most intrinsic mechanism for explaining the observed
baryon asymmetry in the current universe.

Up to this point, a range of scenarios for leptogenesis have been put forth,
contingent on the methods of producing the heavy Majorana neutrinos $N_{i}$ 
\cite{U4,U5,U6,U7}. Of these scenarios, the most widely recognized is the
conventional approach involving the thermal generation of $N_{i}$ during the
early phases of the universe. A comprehensive analysis \cite{U4} reveals
that a satisfactory lepton asymmetry, capable of accounting for the existing
baryon asymmetry can be achieved when the reheating temperature $T_{re}$
following inflation is approximately in the range of ${\mathcal{O}}(10^{10})$
$GeV$. However, delving into leptogenesis within the context of supergravity
introduces a cosmological issue concerning gravitinos: the production of an
excessive number of gravitinos occurs with a $T_{re}$ around $10^{10}$ $GeV$%
, to maintain the viability of big-bang nucleosynthesis (BBN) \cite{U8,U9}.
To tackle this predicament, one approach is to postulate that the gravitino
serves as the lightest supersymmetry (SUSY) particle with a mass denoted as $%
m_{3/2}$, which falls within the range of around $\left( 10-100\right) $ $%
GeV $ \cite{U10,U11,U12,U13}.

In this work, we calculate the gravitational wave spectrum projected by the
cosmological breaking of B-L symmetry. This spectrum is influenced by
various sources as enumerated earlier, including the inflationary period.
Our primary focus lies in the distinctive traits of the GW spectrum
attributed to binary systems of black holes originating in the early
universe. Additionally, we aim to establish a relationship within the GW
spectrum to the values at the phase transition of the hybrid inflation\
which can help identify the model parameters linked to specific
characteristics in the spectrum.

Confirming the inflationary paradigm is strongly associated with the
identification of relic gravitational waves (GWs), which are considered a
clear and unambiguous signal. In recent studies, there has been significant
progress in observational cosmology. Various collaborations, such as
NANOGrav\ \cite{B11,B12}, European PTA (EPTA)/Indian PTA (InPTA) \cite{W1,W2}%
, and Parkes PTA (PPTA) \cite{W3}, have presented compelling evidence
supporting the existence of a Stochastic Gravitational Wave Background
(SGWB) in the frequency range of nanohertz (nHz). While these findings are
primarily attributed to astrophysical sources, it remains crucial to
investigate their potential cosmological origins.

The origin of supermassive black holes (SMBHs) remains uncertain, yet
numerous hypotheses have been put forth. The correlation noted earlier
between the masses of black holes and the dispersion of stellar velocities
can be interpreted in diverse manners, ultimately leading to the question of
which came first: the host galaxy or the SMBHs \cite{V1,V6}. Thanks to the
impressive capabilities of the James Webb Space Telescope (JWST), we now
have the opportunity to gain a more profound understanding of the conclusion
of the cosmic dark ages. To date, the JWST has successfully pinpointed
numerous promising galaxy candidates situated at remarkably high redshifts 
\cite{V7,V8,V9}. Interestingly, within the JWST Cosmic Evolution Early
Release Science (CEERS) program, a collection of massive galaxy candidates
situated in the redshift range of $6.5\lesssim z\lesssim 9.1$ has been
detected. These candidates exhibit estimated stellar masses $M_{\ast
}\gtrsim 109M_{\odot }$ (where $M_{\odot }=1.99\times 10^{30}$ $kg$
represents the solar mass). Notably, among these candidates, there exist six
sources possessing stellar masses $M_{\ast }\gtrsim 10^{10}M_{\odot }$
within the redshift range of $7.4\lesssim z\lesssim 9.1$ \cite{V10}. The
investigation of primordial black holes has its roots in research conducted
over five decades ago \cite{V11,V12}, and in recent years, it has garnered
growing attention. There are diverse underlying reasons for this renewed
interest. One example is the possibility that the merging of binary PBHs
could produce gravitational waves that are detectable by collaborations like
LIGO/Virgo/KAGRA \cite{V13,V14}. Primordial black holes represent a
compelling and inherent contender for the role of dark matter (DM) within
specific mass intervals \cite{V15}. In broad terms, there exist two primary
mechanisms through which PBHs can influence the formation of cosmic
structures: the Poisson effect \cite{V16} and the seed effect \cite{V17,V18}%
. When PBHs constitute a substantial proportion of the overall dark matter
content, the Poisson effect becomes prevalent across all scales. Conversely,
if PBHs account for only a minor fraction of the dark matter, the seed
effect dominates on smaller scales.

Observations of quasars around redshift $z\sim 6$ powered by supermassive
black holes (SMBHs, with masses around $10^{8}M_{\odot }$ - $10^{10}M_{\odot
}$) challenge our current understanding of early black hole formation and
evolution. The JWST has revolutionized this field by enabling the study of
massive black holes (MBHs, with masses around $10^{6}M_{\odot }$ - $%
10^{7}M_{\odot }$) up to redshift $z\sim 11$, thus bridging the properties
of these high-redshift quasars to their progenitors. Moreover, Recent claims
of a vigorously accreting massive black hole (MBH) in GN-z11, a well-known
star-forming galaxy at $z\sim 10.6$,  suggest it may be an active galactic
nucleus (AGN). Our work in this paper aims to constrain the accretion speed
and the Eddington ratio of PBHs to explain the existence of galaxies powered
by these MBHs or SMBHs, thereby addressing the gap from early cosmology to
current observations.

This article is organized as follows, in Sec. \ref{sec2} we discuss the
relation of Leptogenesis to the B-L symmetry-breaking mechanism, in Sec. \ref%
{sec3}\ we introduce the physics of gravitational waves spectrum and the
relation to the PTA signals in light of the B-L breaking, in Sec. \ref{sec4}
we Match the PTA signals with gravitational waves emitted by BBH Mergers and
PBH mass predicted by JWST, in Sec. \ref{sec5} we put constraints on rapid
accretion rates of PBHs to explain JWST observations AGNs and SMBHs. The
last section \ref{sec6}\ is devoted to a conclusion.

\section{ Leptogenesis from B-L symmetry breaking}

\label{sec2}

To illustrate the dynamics of Leptogenesis, we need to examine the B-L
symmetry breaking \cite{A1,A2}, the presence of baryon asymmetry in the
current universe suggests that B-L violating processes had a significant
impact on creating the baryon asymmetry during the early universe. The
lepton number operators violating B-L symmetry originate from the
interaction involving massive Majorana neutrinos $N_{i}$\ $\left(
i=1,2,3\right) \ $\cite{A3,A4}, when $C$ and $CP$ are not conserved, the
decays of these massive neutrinos $N_{i}$ naturally give rise to B-L
asymmetry. The scalar potential governing the inflaton field $\varphi $ is
is given by%
\begin{equation}
V\left( \varphi \right) =V_{0}+V_{1l}\left( \varphi /\varphi _{c}\right) ,
\end{equation}%
with $V_{0}=\frac{\lambda }{4}\upsilon _{B-L}^{4}$ and $\upsilon _{B-L}\
=\varphi _{c}$ is the critical value of the hybrid inflation potential, for $%
\varphi \gg \varphi _{c}$%
\begin{equation}
V_{1l}\left( \varphi /\varphi _{c}\right) \simeq \frac{\lambda }{32\pi ^{2}}%
\upsilon _{B-L}^{4}.
\end{equation}%
When $\varphi $ falls below $\varphi _{c}$, the B-L Higgs boson undergoes a
tachyonic instability, meaning it acquires a negative mass squared. This
sudden occurrence leads to the inflation coming to an end and the
spontaneous breaking of B-L. The slow-roll parameters $\epsilon $ and $\eta $
associated with hybrid inflation are given as follows%
\begin{eqnarray}
\epsilon &\approx &\frac{\lambda }{16\pi ^{2}}\left\vert \eta \right\vert ,
\\
\eta &\approx &-\frac{\lambda M_{p}}{32\pi ^{3}\varphi _{k}^{2}}\approx -%
\frac{1}{2N_{k}}
\end{eqnarray}

\begin{figure}[tbp]
\resizebox{0.6\textwidth}{!}{  \includegraphics{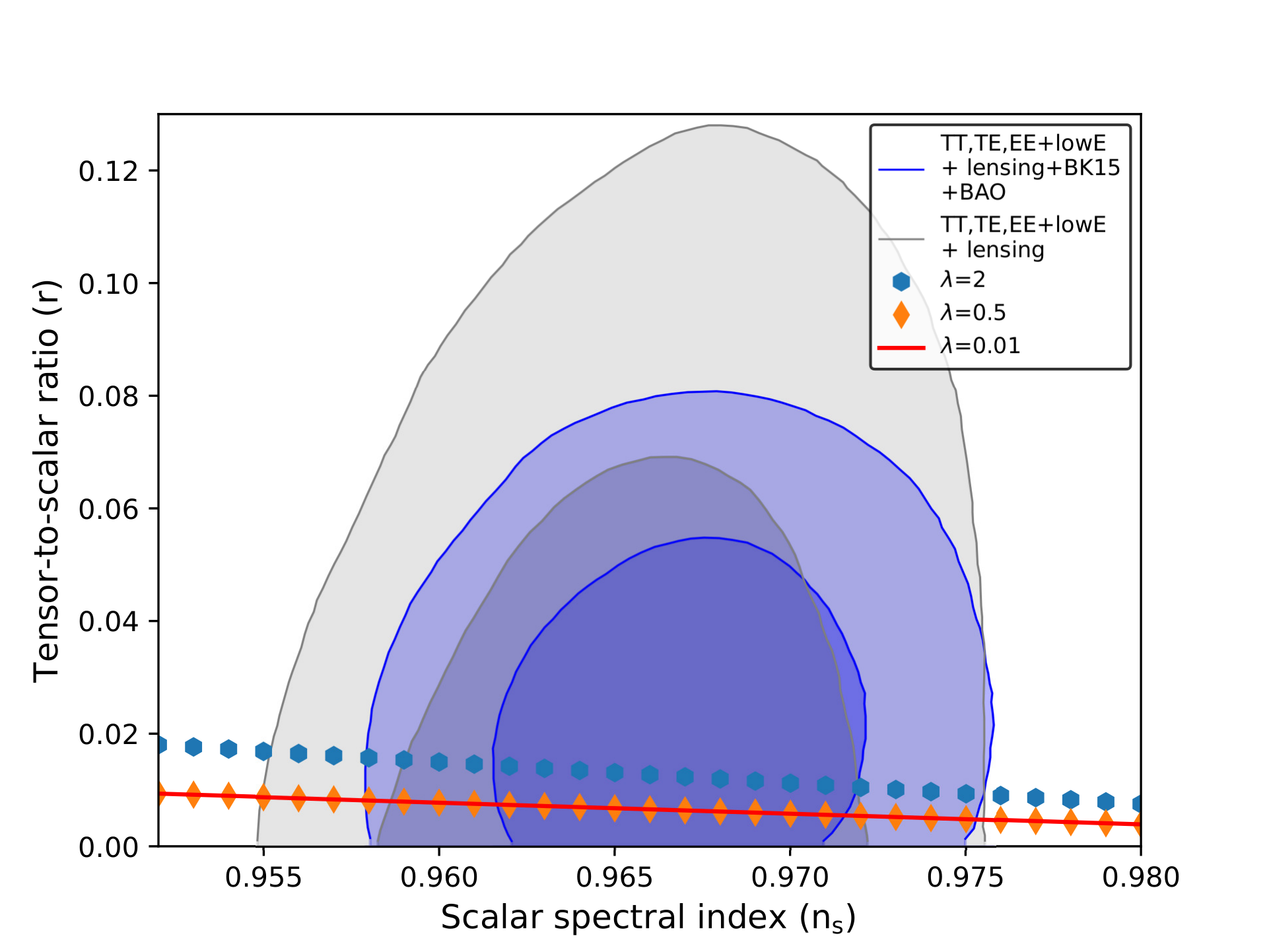}
} 
\caption{$r$ as a function of $n_{s}$ for different values of $\protect%
\lambda $}
\label{fig:1}
\end{figure}

Considering our current study, Fig. \ref{fig:1} shows that the
tensor-to-scalar ratio, $r$, shows good consistency with the predicted
values of $r$ and decreases slowly with respect to the spectral index, $%
n_{s} $. The results show good consistency for a wide range of model
parameters with the latest observations from Planck data. Moreover, the
parameter $\lambda $\ provide good consistency for values bounded as $%
\lambda \leq 0.5$\ following the observational data constraints on $r$ and $%
n_{s}$.

The scalar perturbations can be characterized by the curvature power
spectrum, which describes the specific mode with a wave number k. When
applying the slow-roll approximation, one obtains an analytical expression
at the point of horizon crossing.%
\begin{equation}
{\mathcal{P}}_{s}(k)\equiv \left. \frac{k^{3}}{2\pi ^{2}}\left\vert {%
\mathcal{R}}_{k}\right\vert ^{2}\right\vert _{k\ll aH}\simeq \left. \frac{1}{%
8\pi ^{2}\epsilon }\frac{H^{2}}{M_{p}^{2}}\right\vert _{k=aH},
\end{equation}%
here $3M_{p}^{2}H^{2}=V_{0}\ $and ${\mathcal{R}}_{k}$\ represents the
Fourier component of the comoving curvature perturbation. Similarly, the
tensor perturbations $h_{\mathbf{k}}^{\lambda }$ can be quantified using the
tensor power spectrum. By employing the slow-roll approximation, it becomes
possible to derive an analytical expression for the tensor power spectrum as
follows 
\begin{equation}
{\mathcal{P}}_{t}(k)\equiv \left. \frac{k^{3}}{2\pi ^{2}}\left\vert h_{%
\mathbf{k}}^{\lambda }\right\vert ^{2}\right\vert _{k\ll aH}\simeq \left. 
\frac{2}{\pi ^{2}}\frac{H^{2}}{M_{p}^{2}}\right\vert _{k=aH}.
\end{equation}

For both previous\ quantities of the power spectrum, the value of ${\mathcal{%
P}}_{s}\equiv \Delta _{s}^{2}$ is measured by the PLANCK satellite, $\Delta
_{s}^{2}\simeq 2.101\times 10^{-9}$ \cite{A44}, while for\ the tensor power
spectrum ${\mathcal{P}}_{t}\equiv \Delta _{t}^{2}=$ $r\Delta _{s}^{2}$ \cite%
{A45}. One should note that, slow-roll approximated solution for the scalar
power spectra enhanced scalar fluctuations, which means that small-scale
fluctuations may become large enough for PBH to be overproduced. On the
other hand, the tensor power spectra can act as a source for gravitational
wave background.

\begin{figure}[tbp]
\resizebox{0.6\textwidth}{!}{  \includegraphics{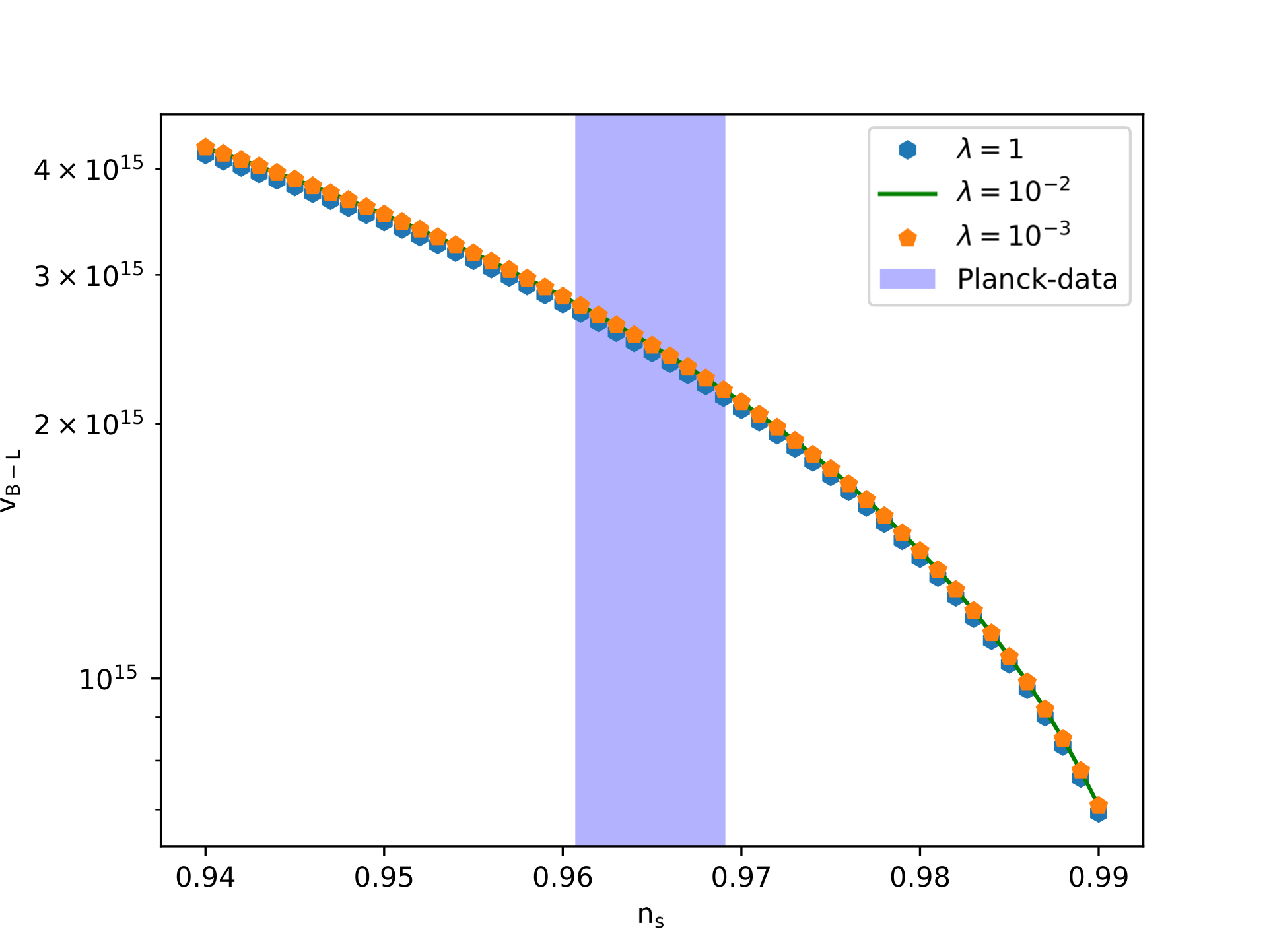}
} 
\caption{Constraints on $\protect\upsilon _{B-L}$ according the recent
Planck's bound on $n_{s}.$}
\label{fig:2}
\end{figure}

Fig. \ref{fig:2} discusses the observational constraints on the critical
value that leads to the B-L symmetry breaking from Planck's data, the
results shows that $\upsilon _{B-L}\propto 10^{15}GeV$ and must be bounded
around $\sim \left[ 2\times 10^{15},5\times 10^{15}\right] GeV,$ our
findings are also consistent with with previous results \cite{A46,A5} where
they proposed that the aim behind $\upsilon _{B-L}$\ value is to adjust its
value to be in proximity to the GUT scale. Specifically, after conducting a
thorough analysis of hybrid inflation, considering the generation of cosmic
strings and non-canonical effects in the Kahler potential, the established
values of $\upsilon _{B-L}$ are consistent with all pertinent observations.

The existence of multiple heavy Majorana neutrino species ($N_{i}$) leads to
two mechanisms that reduce the current lepton asymmetry. In previous
findings (Ref. \cite{A5}) they considered a normal hierarchy of the
right-handed neutrinos, where the most significant contribution to these
processes come from the lightest right-handed neutrino ($N_{1}$), which
leads to the fact that the time evolution of the reheating process is
primarily governed by two key factors: $\Gamma _{S}^{0}$ and $\Gamma
_{N_{1}}^{0}$. These quantities represent the vacuum decay rate of the B-L
Higgs bosons and their superpartners, as well as the vacuum decay rate of
the heavy (s)neutrinos belonging to the first-generation \cite{A5}.%
\begin{eqnarray}
\Gamma _{S}^{0} &=&\frac{m_{S}}{32}\left( \frac{M_{1}}{\upsilon _{B-L}}%
\right) ^{2}\left[ 1-\left( \frac{2M_{1}}{m_{S}}\right) ^{2}\right] ^{1/2} \\
\Gamma _{N_{1}}^{0} &=&\frac{\tilde{m}_{1}}{4\pi }\left( \frac{M_{1}}{%
\upsilon _{u}}\right) ^{2}
\end{eqnarray}

In the Froggatt-Nielsen flavor model, which serves as the basis for our
previous study in Ref. \cite{A6}, the masses of Higgs and (s)neutrinos,
denoted as $m_{S}$, $M_{1}$, and $\tilde{m}_{1}$, are expected to take
specific values: $m_{S}=1.6\times 10^{13}GeV$, $M_{1}=5.4\times 10^{10}GeV$
and $\tilde{m}_{1}=4.0\times 10^{-2}eV$. The number of free and independent
parameters in this model, represented by the two neutrino masses M1 and m1,
correspond exactly to the two decay rates $\Gamma _{S}^{0}$ and $\Gamma
_{N_{1}}^{0}$. The study was conducted in Refs.\ \cite{A2,A5,A6} aimed to
gain a detailed and time-resolved understanding of the reheating process. A
remarkable finding from this research is the presence of an approximate
plateau in the radiation temperature during reheating after the B-L phase
transition. This plateau occurs around the time when the heavy (s)neutrinos
decay. The constancy of the temperature over an extended period is a direct
result of a temporary balance between entropy production and cosmic
expansion. The temperature at which this plateau occurs serves as a
characteristic temperature scale for both leptogenesis and the thermal
production of gravitinos. The next section will be devoted to studying the
impact of B-L symmetry breaking on the gravitational wave background.

\section{ Gravitational Waves Background}

\label{sec3}

\subsection{Primordial Gravitational Waves}

Detecting the background of primordial gravitational waves would serve as
compelling evidence to support the inflation paradigm, shedding light on the
fundamental physics of the early Universe. In this direction, recent
findings have been dedicated to studying the Primordial Gravitational Waves
(PGW) at the first stages following inflation \cite{K1,K2,K3,K4,K5,K6,K7,K8}%
. In the case of the inflationary scenario, the occurrence of a kinetic
epoch preceding inflation results in a distinctive blue tilt in the
primordial gravitational wave spectra at higher frequency ranges \cite%
{B1,B2,B3,B4}. Gravitational waves are characterized as the
transverse-trace-less component of the metric perturbation. In linear order
perturbation theory, there is no coupling between scalar, vector, and tensor
modes. Thus, for primary gravitational waves in a spatially flat FLRW
background, we can express the metric element as follows:%
\begin{equation}
ds^{2}=-dt^{2}+a^{2}(t)(\delta _{\mu \nu }+h_{\mu \nu })dx^{\mu }dx^{\nu },
\end{equation}%
given that the tensor mode fulfills $\left( h_{\mu \mu }=h_{00}=\partial
^{\mu }h_{\mu \mu }=0\right) .$\ $h_{\mu \nu }\left( t,x\right) $ can be
decomposed into its Fourier mode, each associated with two polarization
tensors which satisfy the equation of motion%
\begin{equation}
h_{\mathbf{k}}^{\lambda \prime \prime }(\eta )+2{\mathcal{H}}h_{\mathbf{k}%
}^{\lambda \prime }(\eta )+k^{2}h_{\mathbf{k}}^{\lambda }(\eta )=0,
\end{equation}%
where $\left( ^{\prime }\right) $ represents the derivative with respect to
conformal time $\eta ,$\ $d\eta =$\ $\frac{dt}{a}$, and ${\mathcal{H}}=\frac{%
a^{\prime }}{a}.$ The normalized gravitational wave energy density spectrum
is defined as the energy density per logarithmic frequency interval.%
\begin{equation}
\Omega _{gw}(k)=\frac{1}{\rho _{c}}\frac{d\rho _{gw}}{d\ln k},
\end{equation}%
where $\rho _{c}$ is the total energy density. Moreover,%
\begin{equation}
\Omega _{gw,0}(k)=\frac{1}{12}\left( \frac{k^{2}}{a_{0}^{2}H_{0}^{2}}\right) 
{\mathcal{P}}_{h}(k),
\end{equation}%
with ${\mathcal{P}}_{h}(k)\equiv \frac{k^{3}}{\pi ^{2}}\sum_{\lambda
}\left\vert h_{\mathbf{k}}^{\lambda }\right\vert ^{2}.$ By considering the
the scale at horizon re-entry $\left( k=a_{hc}H_{hc}\right) $ and the Hubble
parameter at different epochs, we can obtain the primary gravitational wave
spectrum at the present time for mode re-entry during the matter $\left(
M\right) $, radiation $\left( R\right) $, and kinetic $\left( K\right) $
eras, respectively as follows:\ 
\begin{eqnarray}
\Omega _{gw,0}^{\left( M\right) } &=&\frac{1}{24}\Omega _{m,0}^{2}\frac{%
a_{0}^{2}H_{0}^{2}}{k^{2}}{\mathcal{P}}_{t}~~~~~~\left( k_{0}<k\leq
k_{eq}\right) , \\
\Omega _{gw,0}^{\left( R\right) } &=&\frac{1}{24}\Omega _{r,0}^{2}\left( 
\frac{g_{\ast }}{g_{\ast 0}}\right) \left( \frac{g_{\ast s}}{g_{\ast s0}}%
\right) {\mathcal{P}}_{t}~~~~~~\left( k_{eq}<k\leq k_{r}\right) , \\
\Omega _{gw,0}^{\left( K\right) } &=&\Omega _{gw,0}^{\left( R\right) }\left( 
\frac{k}{k_{r}}\right) ~~~~~\left( k_{r}<k\leq k_{\max }\right) ,
\end{eqnarray}

\begin{figure*}[h]
\resizebox{1\textwidth}{!}{  \includegraphics{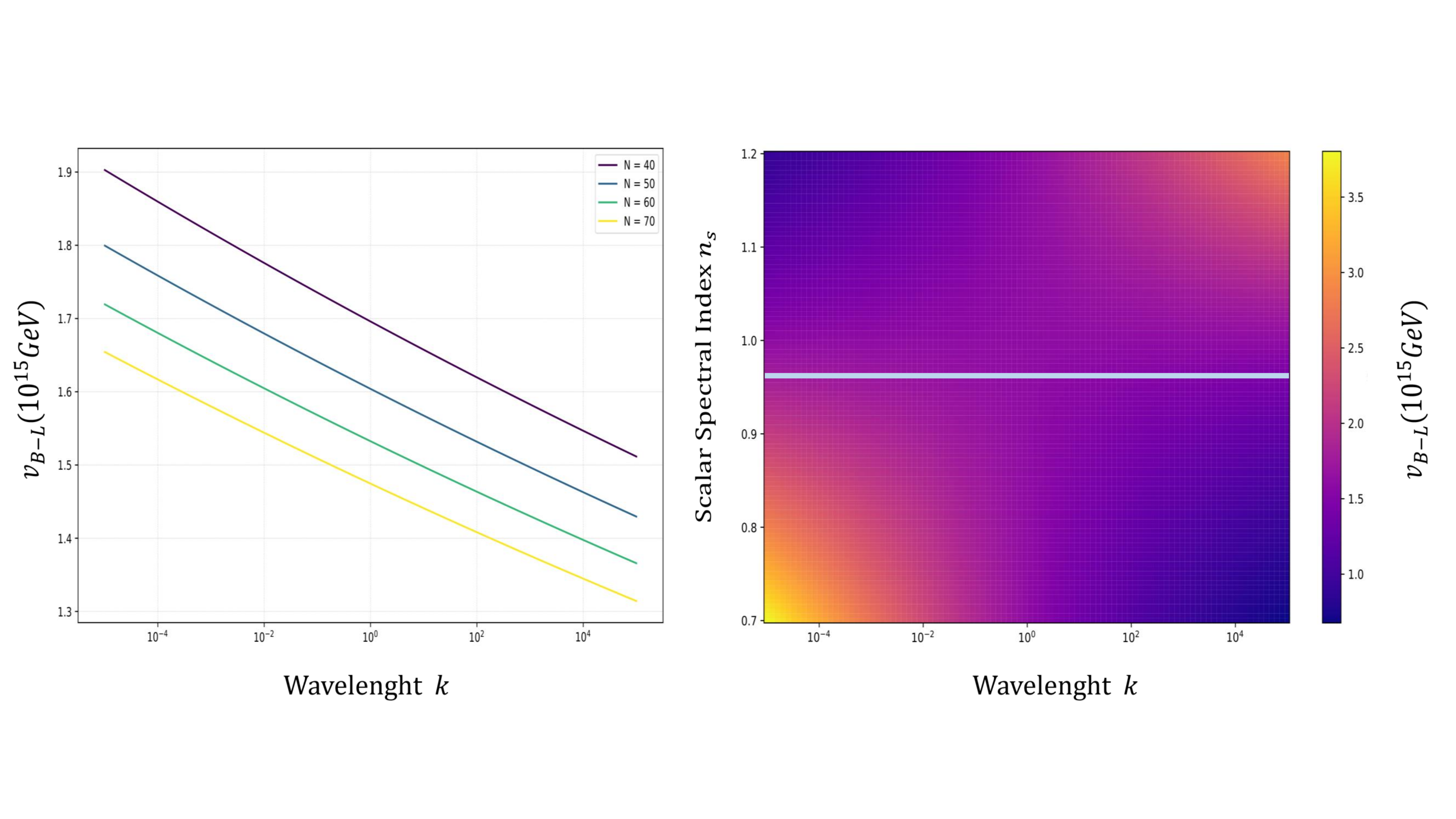}
}
\caption{ The evolution of the B-L asymmetry parameter considering the
variation of the scale $k$ and the spectral index $n_{s}$. }
\label{fig:3}
\end{figure*}

Before we go through more details on the gravitational wave density
Considering the effect of the critical value $\upsilon _{B-L}$ and the scale 
$k.$ In Fig. \ref{fig:3} we study the evolution of the B-L symmetry breaking
parameter $\upsilon _{B-L}$ as a function of the wavelength $k$\ considering
the index spectral observational bound effects on $\upsilon _{B-L}.$ The
results are found considering the form of the tensor power spectrum from the
previous section and the fact that ${\mathcal{P}}_{t}\propto k^{n_{s}-1}$,\
we conclude that considering superhorizon scales the potential parameter
will decrease as we take higher values of inflation e-folds number.
Additionally, the spectral index shows that for both smaller and larger
scales the $\upsilon _{B-L}$\ is best considered around $2\times
10^{15}GeV\leq \upsilon _{B-L}<3\times 10^{15}GeV.$

\begin{figure*}[h]
\resizebox{1.1\textwidth}{!}{  \includegraphics{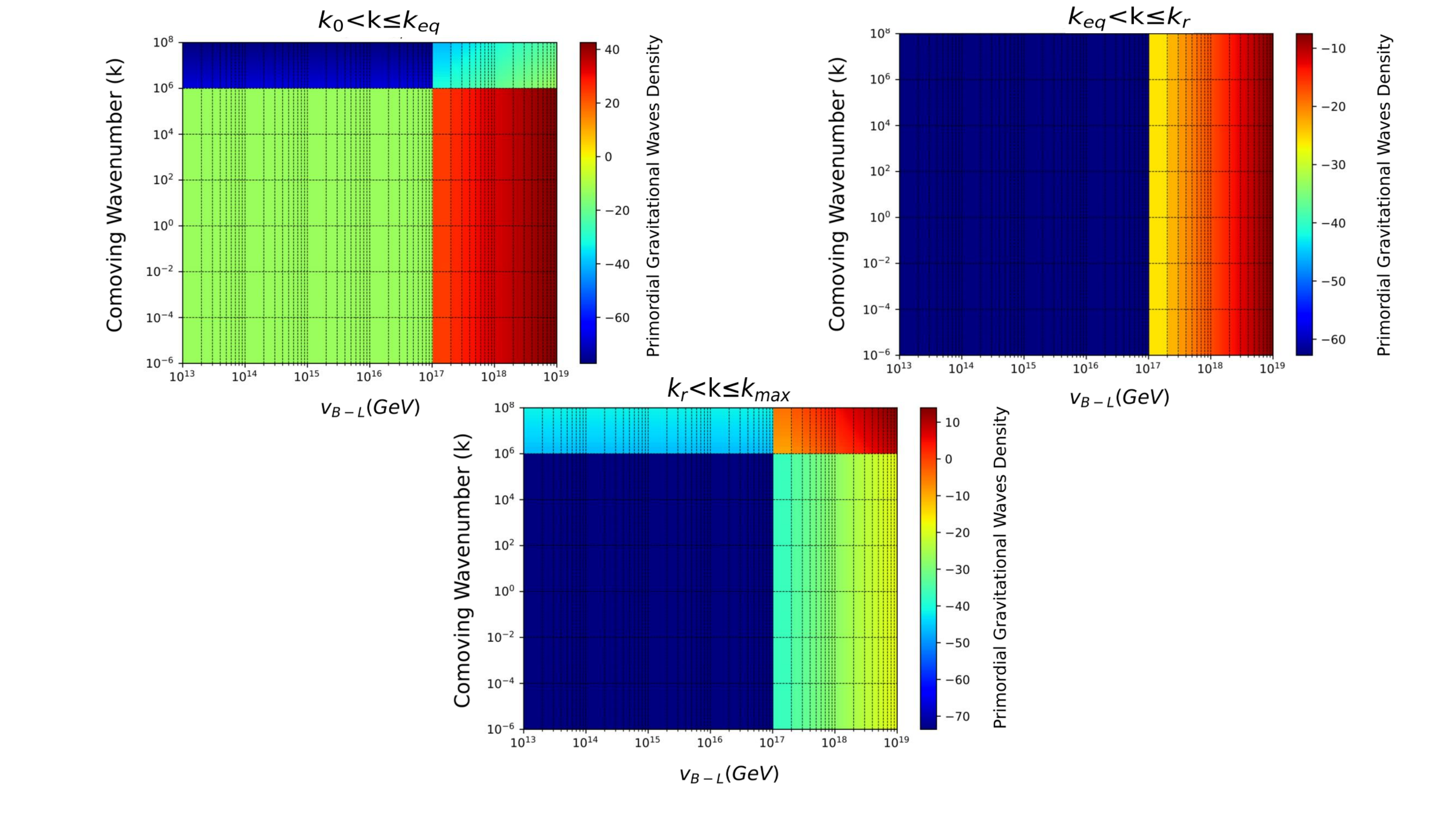}
}
\caption{ Primordial gravitational wave evolution with respect to the scale $%
k$\ and $\protect\upsilon _{B-L}$. }
\label{fig:4}
\end{figure*}

Fig. \ref{fig:4} represents the variation of the density of primordial
gravitational waves with respect to the scale $k$ and the $\upsilon _{B-L}$
parameter. The color scale on the right of the plots represents the
logarithm of the density of PGWs. Regions with higher color intensity
(warmer colors) indicate higher density, while regions with lower color
intensity (cooler colors) indicate lower density. The change in the critical
value $\upsilon _{B-L}$ that corresponds to B-L asymmetry has proportional
effects on $\Omega _{gw,0}$ for all matter, radiation, and kinetic eras. In
fact, as we go higher in $\upsilon _{B-L}$ values specifically around $%
\upsilon _{B-L}\propto 10^{17}GeV$ \ we notice a significant transition in
the density of PGWs. Moreover, the transition of $\Omega _{gw,0}$
corresponds to larger scales (e.g., superhorizon scales) given by $k\propto
10^{6}$\ which are noticed only for the matter and kinetic eras. The peak
region in the heat map, where the color intensity is highest, indicates the
preferred combination of $\upsilon _{B-L}$ and $k$ that leads to the highest
density of primordial gravitational waves. This combination represents the
characteristic scale of the gravitational waves generated during inflation.
To conclude, we can see that the proposed scenario of Leptogenesis has great
implications for the observed values of PGWs density which can be predicted
by PTA\ signals if made fine tuning to the values of the scale $k$ and $%
\upsilon _{B-L}$\ for the matter, radiation, and kinetic eras. Next, we will
see how much can the NanogGrav predictions reproduce the theoretical values
predicted by the chosen model of the density of PGWs.

\subsection{Matching Gravitational Waves to the PTA Signals}

The time evolution of a gravitational wave field, represented by $h_{\mathbf{%
k}}(\eta _{i})$ at an initial conformal time $\eta _{i}$, and characterized
by its tensor spectrum, can be obtained by determining the GW transfer
function ${\mathcal{T}}\mathbf{(\eta ,k)=}h_{\mathbf{k}}(\eta )/h_{\mathbf{k}%
}(\eta _{i})$ \cite{B5}. Here, $h_{\mathbf{k}}(\eta )$ is evaluated at a
conformal time $\eta \gg \eta _{i}$ \cite{B6,B7}. The relevant quantity for
GW direct detection experiments, such as Pulsar Timing Arrays, are the
spectral energy density parameter $\Omega _{gw}(k)$ \cite{B8}:

\begin{equation}
\Omega _{gw}(k)\equiv \frac{k^{2}}{12H_{0}}{\mathcal{T}}\mathbf{(\eta }_{0}%
\mathbf{,k)}{\mathcal{P}}_{t}(k),
\end{equation}%
using $\mathbf{\eta }_{0}$ to represent the present conformal time and $%
H_{0} $ to denote the Hubble constant \cite{B9}. For the purpose of this
investigation, we focus on $\Omega _{gw}(k)$ at PTA scales, where $f\sim {%
\mathcal{O}}(10-9)Hz$. The corresponding wavenumbers $k\sim {\mathcal{O}}%
(10^{6})Mpc^{-1}$ are much larger than the wavenumber associated with a mode
crossing the horizon at matter-radiation equality. In other words, the modes
observed on PTA scales crossed the horizon deep in the radiation era, well
before the matter-radiation equality. In the regime where $k\gg k_{eq}$, the
GW spectral energy density associated with PTA signals can be expressed as 
\cite{B10,B11,B12,B13,B14,B141}:%
\begin{equation}
\Omega _{gw}(f)=\frac{2\pi ^{2}}{3H_{0}^{2}}A^{2}\frac{f^{5-\gamma }}{%
f_{yr}^{\gamma -3}}.
\end{equation}

The anticipated signal of the Stochastic Gravitational Wave Background
(SGWB) generated by merging Black Hole Binaries (BHBs) mergers corresponds
to $\gamma =13/3$ and $\alpha \simeq 5-\gamma .$ The relationship linking
the amplitude of the Pulsar Timing Array (PTA) signal, denoted as $A$, with
the cosmological parameters were found to be \cite{B8,B15}%
\begin{equation}
A=\sqrt{\frac{45\Omega _{m}^{2}A_{s}}{32\pi ^{2}(\eta _{0}k_{eq})^{2}}}c%
\frac{f_{yr}}{\eta _{0}}\left( \frac{f_{yr}^{-1}}{f_{\star }}\right) ^{\frac{%
n_{T}}{2}}\sqrt{r}.
\end{equation}

The dependence on $n_{T}$ in this context arises from the substantial "lever
arm" between the Cosmic Microwave Background (CMB) pivot frequency, where As
is constrained, and the frequency of the Pulsar Timing Array (PTA) signal $%
\left[ 1yr^{-1}\right] $.

\begin{table*}[tbp]
\centering%
\begin{tabular}{c|c|c|c||c|c}
\hline
$\lambda $ & $v_{B-L}(GeV)$ & $r$ & $A$ & $f(Hz)$ & $h^{2}\Omega _{gw}(f)$
\\ \hline\hline
$1\times 10^{-5}$ & $2\times 10^{15}$ & $0.0174$ & $6.02\times 10^{-15}$ & $%
4\times 10^{-8}$ & $5.06\times 10^{-8}$ \\ \hline
$1.5\times 10^{-5}$ & $2\times 10^{15}$ & $0.0261$ & $7.37\times 10^{-15}$ & 
$2\times 10^{-8}$ & $5.36\times 10^{-8}$ \\ \hline
$2\times 10^{-5}$ & $2.1\times 10^{15}$ & $0.0423$ & $9.38\times 10^{-15}$ & 
$1\times 10^{-8}$ & $6.14\times 10^{-8}$ \\ \hline
$2\times 10^{-5}$ & $2.2\times 10^{15}$ & $0.0510$ & $1.03\times 10^{-14}$ & 
$8\times 10^{-7}$ & $6.62\times 10^{-7}$ \\ \hline
$2.5\times 10^{-5}$ & $2.3\times 10^{15}$ & $0.0762$ & $1.26\times 10^{-14}$
& $6\times 10^{-7}$ & $8.59\times 10^{-7}$ \\ \hline
$3\times 10^{-5}$ & $2.4\times 10^{15}$ & $0.1084$ & $1.50\times 10^{-14}$ & 
$5\times 10^{-7}$ & $1.11\times 10^{-7}$ \\ \hline
\end{tabular}%
\caption{Testing the density of gravitational waves predicted by PTA as
functions of several parameters related to the potential considered for
describing the Leptogenesis mechanism, here we've inserted the best-fit
values for the cosmological parameters as per the Planck results 
\protect\cite{A44}, $\Omega _{m}=0.315,$\ $\protect\eta _{0}\simeq 1.38,$\ $%
A_{s}=2,1\times 10^{-9},k_{eq}\simeq 0.01,f_{\star }\approx 7.7\times
10^{-17}Hz,$\ $f_{yr}\simeq 3.1\times 10^{-8}Hz$ and $H_{0}\equiv
100~h~km/s/Mpc.$\ \ \ \ \ \ \ \ \ }
\label{table:1}
\end{table*}

Table \ref{table:1} corresponds to the NanoGrav proposed model of the
density $h^{2}\Omega _{gw}(f)$ as a function of the frequency $f,$
scalar-to-tensor ratio $r$, the amplitude\ of PTA signal that is related to
the proposed potential parameters that explain the Leptogenesis phenomena.
The model for $\Omega _{gw}(f)$ depends on various cosmological parameters
and other constants defined previously. We compute $h^{2}\Omega _{gw}(f)$ is
an analytical expression based on the theoretical model for primordial
gravitational waves proposed recently, $h^{2}\Omega _{gw}(f)$ is the the
fractional energy density of gravitational waves in the universe. It
represents the amount of energy carried by gravitational wave background
relative to the critical energy density required to achieve a flat universe.
By looking at the table identify the right parameters that can predict the
higher or lower density of gravitational waves $h^{2}\Omega _{gw}(f)$ which
reproduce the NANOGrav predictions, one can clearly see that for values of $%
r\lesssim 0.06$, with the spectrum is blue $\left( n_{T}>0\right) $ which we
consider to take $n_{T}=1.1$, the amplitude of PTA scales will be $A\lesssim
10^{-20}$ which will provide good predictions for $h^{2}\Omega _{gw}(f)$
values. Moreover, the table provides insights into the behavior of GWB
density with respect to frequency, the scalar-to-tensor ratio, and the B-L
asymmetry parameter, we can conclude that the results show good consistency
with the predicted bounds on the density and the frequency once we fine-tune
the inflationary parameter with parameters associated with the PTA signals.

\section{Matching PTA Signals with Gravitational Waves Emitted by BBH
Mergers and the required PBH mass to explain the JWST Data}

\label{sec4}

\subsection{Gravitational waves from BBH mergers and PTA Interpretation}

In the early universe, supermassive primordial black holes with masses
around ${\mathcal{O}}(10^{18}M_{\odot })$ could potentially emerge through
various mechanisms, particularly during the radiation-dominated era.
Nevertheless, before matter-radiation equality and as a consequence of the
universe's expansion, multiple primordial black holes (PBHs) have the
opportunity to coexist within the same Hubble patch, allowing for the
formation of isolated pairs of two PBHs \cite{C1}. PBH mass can be
calculated as follows \cite{C2},%
\begin{equation}
M_{PBH}\simeq M_{\odot }\left( 5\times 10^{12}/\left( 1+z_{f}\right) \right)
^{2}
\end{equation}

In principle, binary primordial black holes (PBHs) can form either before 
\cite{C3}, or after \cite{C4,C5,C6} matter-radiation equality. In the
context of the binary black hole (BBH) merger, our focus will be on
examining the observable implications of this merger rate, which is
represented by a power law distribution model in the following form \cite{C7}%
:%
\begin{equation}
\frac{dN_{merg}}{dtdV}\equiv {\mathcal{R}}\left( z\right) =R_{0}\left(
1+z\right) ^{\alpha },
\end{equation}%
for a current merger rate of $R_{0}=9-35Gpc^{-3}yr^{-1}$, the parameter $%
\alpha $ is determined to be $2.7_{-1.9}^{+1.8}$. The total energy released
in gravitational waves, accounting for the inspiral, merging, and ringdown
phases is expressed as follows \cite{C10,C11}:%
\begin{eqnarray}
\frac{dE_{gw}}{df} &=&\frac{\pi ^{2/3}M_{c}^{5/3}}{3} \times \\
&&\left\{ 
\begin{array}{c}
f_{s}^{-1/3}~~~~~~~~~~~~~~~~~~~~~~~~~~~for~~f<f_{1} \\ 
f_{1}^{-1}f^{2/3}~~~~~~~~~~~~~~~~~~for~~f_{1}\leq f<f_{2} \\ 
f_{2}^{-3/4}f_{1}^{-1}\frac{\sigma ^{4}f}{\left( \sigma ^{2}+4\left(
f-f_{2}\right) ^{2}\right) ^{2}}~~~~~for~~f_{2}\leq f<f_{3}%
\end{array}%
\right.  \notag
\end{eqnarray}%
the GW energy spectrum of the merger, denoted as $dE_{gw}/df_{s}$, is
associated with the frequency in the source frame, represented by $f_{s}$.
This frequency in the source frame is related to the observed frequency, $f$%
, through the expression $f=f_{obs}\left( 1+z\right) $. The function $%
E\left( z\right) \equiv H(z)/H_{0}=\left[ \Omega _{r}\left( 1+z\right)
^{4}+\Omega _{m}\left( 1+z\right) ^{3}+\Omega _{\Lambda }\right] ^{1/2}$,
where $M_{c}$ is the chirp mass, $M_{c}=M_{PBH}/2^{1/5}$ \cite{C1}.
Analyzing the spectral attributes of the BBH background typically involves
describing the energy density parameter to characterize the SGWB \cite%
{C12,C13}:

\begin{equation}
\left. \Omega _{gw}(f)\right\vert _{f_{obs}}=\int_{0}^{z_{\sup }}\left.
\Theta _{\nu }(f,z)\right\vert _{f_{obs}}\frac{d{\mathcal{R}}\left( z\right) 
}{dz}dz,
\end{equation}%
here $d{\mathcal{R}}\left( z\right) /dz$\ is the expression for the
differential gravitational wave event rate, while $\Theta _{\nu }$
represents the energy flux emitted per unit frequency by a source located at
a luminosity distance $d_{L}(z)$:%
\begin{equation}
\left. \Theta _{\nu }(f,z)\right\vert _{f_{obs}}=\frac{1}{4\pi d_{L}(z)^{2}}%
\frac{dE_{gw}}{df}\left( 1+z\right) ,
\end{equation}%
Ultimately, it proves beneficial to represent the model in the following
form \cite{C13}%
\begin{eqnarray}
\left. \Omega _{gw}(f)\right\vert _{f_{obs}} &\simeq &4.2\times
10^{-10}\left( \frac{r_{0}}{3.1\times 10^{-2}Mpc^{-3}Myr^{-1}}\right)  \notag
\\
&&\times \left( \frac{M_{c}}{8.7M_{\odot }}\right) ^{\frac{5}{3}}\left( 
\frac{_{f_{obs}}}{100Hz}\right) ^{\frac{2}{3}},
\end{eqnarray}%
As our focus is on the current energy density of gravitational waves, it
becomes necessary to transform the earlier GW spectrum into a contemporary
physical parameters. One can obtain the present GW spectra as functions of
the observable density of GWs which can be formulated as follows:%
\begin{equation}
\Omega _{gw,0}(f)h^{2}=\left. \Omega _{gw}(f)\right\vert _{f_{obs}}\left( 
\frac{g_{\ast j}}{g_{\ast 0}}\right) ^{-\frac{1}{3}}\Omega _{r,0}h^{2}
\end{equation}%
here $h$ is the present dimensionless Hubble constant and $\Omega
_{r,0}h^{2}\simeq 4.15\times 10^{-5}$ is the abundance of radiation today,
we choose $g_{\ast j}/g_{\ast 0}=106.75/3.36\simeq 31.$ Because
gravitational waves (GWs) propagate at the speed of light, they are present
in the Universe similar to radiation and add to the overall radiation
energy. As a result, measurements of the total radiation density from the
Cosmic Microwave Background (CMB) and Big Bang Nucleosynthesis (BBN) for
species beyond the standard model can impose limitations on the overall
energy density attributed to gravitational waves.

\begin{figure*}[h]
\resizebox{1\textwidth}{!}{  \includegraphics{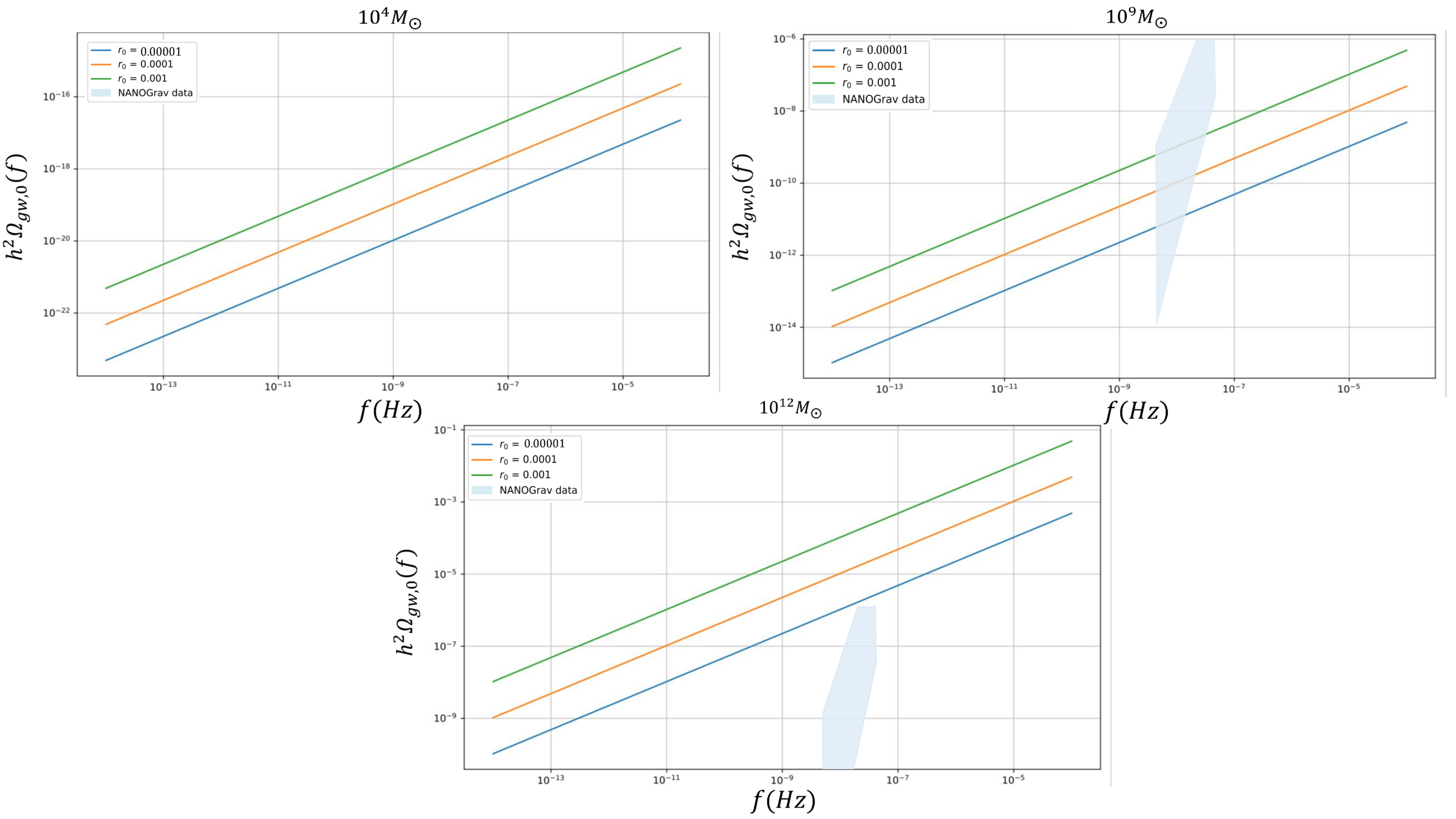}}
\caption{ Primordial gravitational waves evolution with respect to the scale
and the frequency $f$, the light blue region represents the NANOGrav data,
and $r_{0}$ \ is represented by three different colors for each value. }
\label{fig:5}
\end{figure*}

In Fig. \ref{fig:5}, we show the abundance of the energy density of
gravitational waves as a function of the present value of the frequency. We
also display three expected values of $r_{0}=10^{-5},10^{-4},$ and $10^{-3}$%
. We numerically compute the energy spectrum of the GWs to test the effect
PBHs mass. The GW spectrum increases the frequencies and spans over the
NANOGrav sensitivity region for $M_{PBH}\propto 10^{9}M_{\odot }$. We
conclude that the BBH system can partially contribute to the explanation of
the energy spectrum at nanohertz frequencies. However, PBHs mass $%
M_{PBH}\geq 10^{12}M_{\odot }$ are not consistent with the expected NANOGrav
GWs observatories. Moreover, the parameter $r_{0}$ of the chosen model can
affect the behavior of GWs density, which means that higher values of $r_{0}$%
\ increase the density $\Omega _{gw,0}(f)h^{2}$. Here, we mention that the
stochastic gravitational waves from PBH binaries are important stochastic
gravitational wave background sources that are used to probe the Universe
and can provide a possible explanation for the recent NANOGrav results and
JWST\ predictions as we will see in the rest of the paper.

\subsection{The abundance of PBH and its dependence on the B-L Breaking
Mechanism}

A process of PBH formation was first discussed in \cite{V2,V3,V4,V5} which
hinges on the dynamics of inflation, involving a scalar field with two
distinct vacuum states. These states become unevenly distributed following
the conclusion of inflation. As the universe transitions out of the
inflationary phase and cools down, the expansion rate decreases.
Consequently, the scalar field is allowed to move freely towards the lowest
points in its potential energy. This scenario gives rise to regions within
the universe where less probable vacuum states exist, surrounded by a more
favored vacuum state. Essentially, this implies that once vacuum states
become populated during the Friedmann-Robertson-Walker epoch, enclosed
regions of less probable vacuum form, resembling islands. Over time, these
closed vacuum walls can become causally connected and subsequently collapse,
resulting in the formation of black holes. The mass of these black holes
corresponds to that of the originating vacuum wall. Moreover, PBH formation
during the early Universe generally results from the collapse of significant
density fluctuations \cite{D1}. Researchers have explored limitations on the
prevalence of PBHs through observations \cite{D2,D3}, suggesting their
abundance to be lower than 10-20 of the overall energy density of the
Universe. A crucial requirement is that the density contrast needs to
surpass the critical threshold of $\delta _{c}=0.414$ \cite{D4}. This
condition ensures the formation of PBHs during the radiation-dominated era,
precisely when fluctuations come within the horizon. The estimation of
primordial black hole (PBH) generation involves evaluating the portion of
energy density, which is intricately linked to the variance of density
perturbations. For this purpose, we employ the power spectrum alongside a
window function, denoted as $W\left( kR\right) $ \cite{D5}, to establish the
variance $\sigma (k)$ with a final form given as \cite{D8},%
\begin{eqnarray}
\sigma ^{2} &=&\int_{0}^{\infty }\left( kR\right) ^{2}{\mathcal{P}}%
_{s}(k)W^{2}(\tilde{k}R)\frac{dk}{k}, \\
\sigma ^{2} &\simeq &\frac{4\left( 1+\omega \right) ^{2}}{\left( 5+3\omega
\right) ^{2}}{\mathcal{P}}_{s}(k).
\end{eqnarray}%
The PBH mass fraction can be obtained in an integral form when we assume the
Gaussian fluctuations are given by

\begin{equation}
\beta \simeq \frac{1}{\sqrt{2\pi }}\frac{\sigma }{\delta _{c}}\exp \left(
-\left( \frac{\delta _{c}}{\sqrt{2}\sigma }\right) ^{2}\right) .
\end{equation}%
The current density parameter of the PBHs can be obtained from the relation 
\cite{D8,D9},%
\begin{equation*}
\beta =7.06\times 10^{-18}\Omega _{PBH}\left( \frac{M_{PBH}}{10^{15}g}%
\right) ^{\frac{1}{2}}.
\end{equation*}

\begin{figure}[tbp]
\resizebox{0.7\textwidth}{!}{  \includegraphics{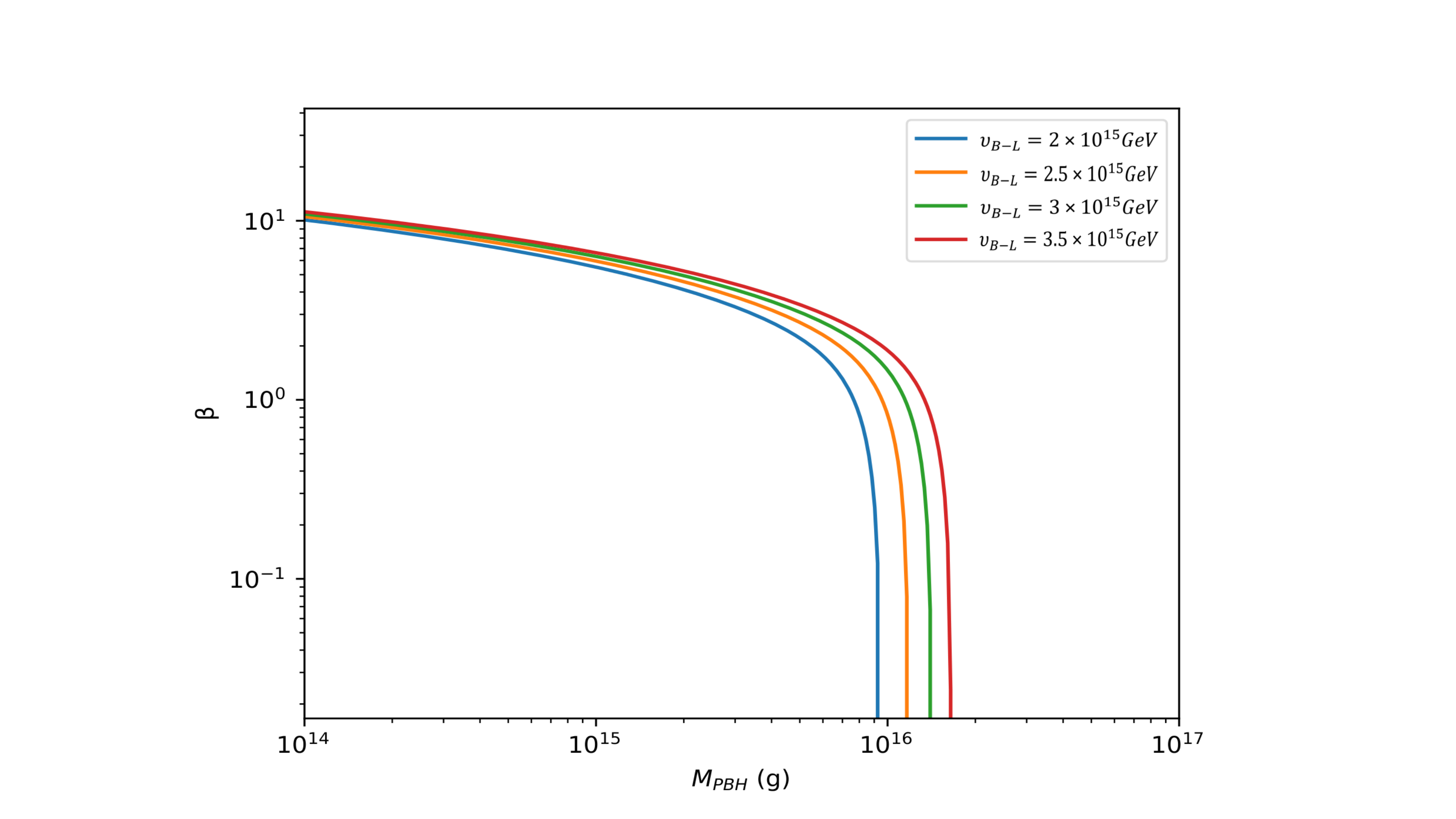}
} 
\caption{The evolution of PBH mass fraction as functions of $M_{PBH}$ for
different values of the potential parameter $v_{B-L}$.}
\label{fig:6}
\end{figure}

Primordial black holes originate from the amplification in the initial power
distribution at smaller scales. Assuming that, during a radiation-dominated
phase in the Universe's evolution, it has achieved the adiabatic limit, and
the curvature perturbation on larger-than-horizon scales becomes immobilized
in comoving terms. Upon the resumption of a scale entering the Hubble
horizon, certain areas could exhibit a significant positive curvature,
analogous to a closed Universe. During this stage, the expansion halts and a
phase of contraction commences. Subsequently, the Hubble-sized area
characterized by substantial positive curvature will undergo a collapse,
leading to the formation of a black hole. The fraction of energy density can
provide an alternative explanation for investigating the collapses of
primordial black holes in the early Universe. Fig. \ref{fig:6} shows the
fraction of the total energy density for spherically symmetric regions
collapsing into PBH as a function of PBH mass for different values of $%
\upsilon _{B-L}$\ parameter, The results show that the curves converge
toward $10^{1}$ where the overproduction regime occurs \ and PBH production
is satisfied, knowing that $\upsilon _{B-L}$ has a notable effect on the
behavior of the fraction of the total energy density where they converge
towards values around $M_{PBH}\propto 10^{16}g$.\ Moreover, $\beta $ shows a
decreasing behavior towards higher values of the fraction $\beta,$ we should
note that lower values $M_{PBH}$\ results a convergence of the fraction of
the total energy density towards a maximal value of $\beta $. It's important
to highlight that in this context, we've neglected the evolution of PBHs
through processes like radiation, accretion, and merging, as these
mechanisms fail to account for present-day observations. Thus, when
contemplating PBH accretion as a means to form massive galaxies, we need
supplementary criteria to elucidate the observations made by the JWST.

\subsection{Investigating the Mass of Primordial Black Holes using JWST
Observations}

In recent studies, there has been significant interest in the investigation
of PBHs, as they have the potential to provide insights into a range of
cosmological occurrences. These include the formation of galaxies with high
redshifts, the characteristics of dark matter, and the origins of
supermassive black holes. After their formation, PBHs have the potential to
undergo changes over cosmic time through the process of accretion. The
accumulation of baryonic matter onto PBHs during this accretion phase can
notably influence their mass \cite{E1,E2}. It's essential to emphasize that
PBHs constitute only a small fraction of the overall dark matter content.
Moreover, each PBH is enveloped by a prevailing dark matter halo, which
grows over time, provided that these PBHs avoid mutual interactions \cite{E3}%
. Sizeable primordial black holes have the potential to act as origins of
fluctuations at a specific mass scale, denoted as MB, either through the
Poisson effect \cite{E4,E5} or the seed effect \cite{E6,E7}. Due to the
presence of a high-density radiation field around each PBH during its
formation, the fluctuation becomes stagnant in the radiation-dominated era.
However, it starts to amplify during the matter-dominated era. The mass $%
M_{B}$, which is bound at the redshift $z_{B}$, is a key factor in this
process \cite{E8}.

\begin{equation}
M_{B}\approx \left\{ 
\begin{array}{c}
\frac{M_{PBH}f_{PBH}}{\left[ \left( 1+z_{B}\right) a_{eq}\right] ^{2}}%
~~\left( Poisson~effect\right) , \\ 
\\ 
\frac{M_{PBH}f_{PBH}}{\left( 1+z_{B}\right) a_{eq}}~~\left(
Seed~effect\right),%
\end{array}%
\right.
\end{equation}%
here $a_{eq}$ is the scale factor at the matter--radiation equality. At this
point we will assume that $M_{B}\sim M_{halo}$ and the galaxy redshift $%
z\sim z_{B}$. Furthermore, we should introduce the star formation efficiency 
$\varepsilon $\ given by,%
\begin{equation}
\varepsilon =\frac{M_{\ast }}{f_{b}M_{halo}},
\end{equation}%
where $M_{halo}$ is the halo mass, and $f_{b}$ is the fraction of baryons in
matter.

\begin{figure*}[h]
\resizebox{1\textwidth}{!}{  \includegraphics{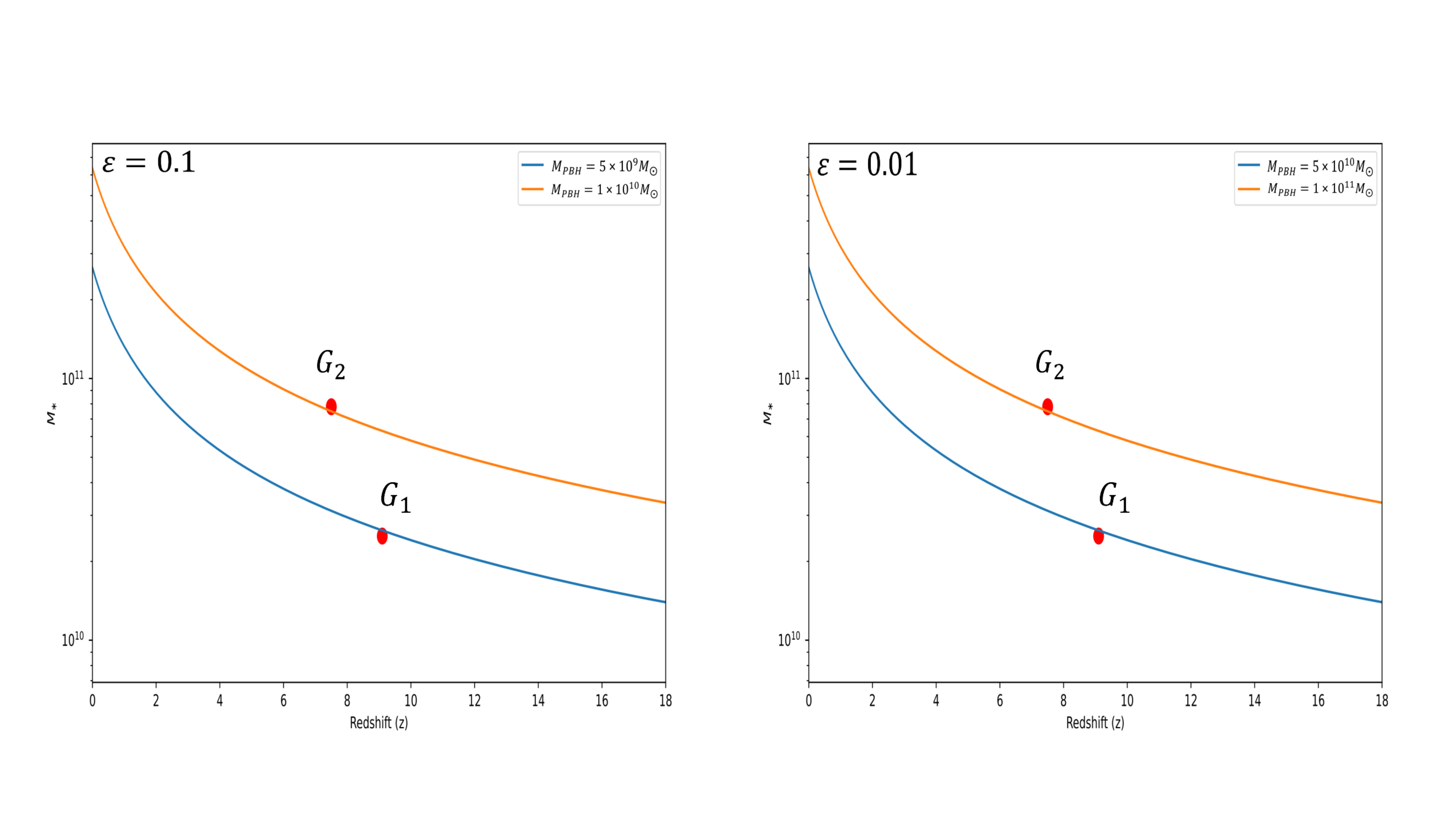}
}
\caption{ The evolution of galaxies stellar masses $M_{\ast }$ as a function
of the redshift for different values of the PBH mass $M_{PBH}$, the two dots
represent the two most massive high-z objects, galaxies 35300 and 38094
referred as $G1$ and $G2$ respectively. }
\label{fig:7}
\end{figure*}

Several investigations have imposed limitations on the mass of PBHs based on
the Poisson effect standpoint. Within this context, PBHs with a mass greater
than $10^{11}M_{\odot }$ have been ruled out by multiple studies due to
their non-detection, with the exception of Phoenix A \cite{E9}. Moreover, in
addition to the restriction posed by the $\mu $-distortion, PBHs with a mass
around $10^{5}M_{\odot }$ and above also face comparatively milder
constraints stemming from sources such as X-ray binaries \cite{E10}, PBH
infall into the Galactic center influenced by dynamical friction \cite{E11},
and statistical properties of the large-scale cosmic structure \cite{E8}.
These combined factors necessitate that the fraction of dark matter composed
of PBHs, $f_{PBH}$ should be in the range of $10^{-4}$ to $10^{-3}$ for PBH
masses spanning from around $10^{5}$ to $10^{11}$ $M_{\odot }$ \cite{E12}.
Additionally, based on data from the high-redshift Lyman-$\alpha $ forest
observations \cite{E13}, primordial black holes with $M_{PBH}f_{PBH}$ $%
\gtrsim $ $170M_{\odot }$ and $f_{PBH}$ $>0.05$ have been excluded. This
indicates a significant unfavorable stance towards the existence of
supermassive PBHs capable of contributing to the Poisson effect. In Fig. \ref%
{fig:7} we have plotted the variation of galaxies' stellar masses $M_{\ast }$
as functions of the redshift $z$, we observe that the stellar mass decreases
slowly with respect to the redshift towards stable values. Moreover, the
galaxy formation efficiency $\varepsilon $ has a notable effect on the PBH
mass $M_{PBH}$\ which could get lower as we take higher values of $%
\varepsilon .$\ Finally, the influence of the seed effect, which operates on
smaller scales, remains a viable explanation for interpreting the
observations made by the JWST observation.

\section{Assessment of JWST Observations: Verifying Rapid Accretion Rates of
PBHs}

\label{sec5}

Recent studies indicate that the farthest quasars identified to date, with
redshifts around $\left[ 6,7.5\right] $, are fueled by supermassive black
holes (SMBHs) weighing approximately $10^{8}M_{\odot }$to $10^{10}M_{\odot }$
\cite{E14,E15,E16,E17,E18,E19,E20}. The presence of these immense black
holes poses a challenge to current theoretical models of black hole
formation and evolution, as critical details such as the nature and mass of
the SMBH seeds, as well as their ability to grow rapidly enough to reach
SMBH status within the first billion years of the Universe, remain
unresolved. Investigating supermassive black holes of around $10^{6}M_{\odot
}$to $10^{7}M_{\odot }$ solar masses at redshifts $z>9$ is crucial for
advancing our understanding in this field \cite{E21,E22}. The recent
assertion of a highly active supermassive black hole in GN-z11, a renowned
galaxy undergoing star formation at $z=10.6$ \cite{E23} aligns with this
pursuit. GN-z11 was initially detected by \cite{E24,E25} and subsequently
examined in the JWST Advanced Deep Extragalactic Survey \cite{E26} using
NIRSpec spectroscopy \cite{E27} and NIRCam imaging \cite{E28}. These
observations indicate that GN-z11 exhibits characteristics consistent with
an Active Galactic Nucleus (AGN), supported by several pieces of evidence
namely the presence of high ionization transitions \cite{E29}. Moreover,
commonly associated with AGNs, as they necessitate high photon energies
typically not generated by stars \cite{E30,E31}.

It's generally assumed that Nuclear Black Holes (BHs) enlarge by drawing in
gas from the nearby environment. This accretion process is governed by both
star formation and mechanical feedback mechanisms, both of which result in a
reduction of gas within the galaxy hosting the BH. According to \cite%
{E32,E33,E34}, it's postulated that nuclear BHs accrete gas according to the
Bondi-Hoyle-Lyttleton (BHL) accretion rate formula:%
\begin{equation}
\dot{M}_{blondi}=\frac{4\pi G^{2}M_{BH}^{2}\left\langle \rho
_{gas}\right\rangle }{\left( \left\langle c_{s}\right\rangle
^{2}+\left\langle \nu _{BH}\right\rangle ^{2}\right) ^{3/2}},
\end{equation}

In the above equation $c_{s}$ is the sound speed, $\nu _{BH}$ is the
velocity of the $BH$ relative to the gas, MBH is the mass of the BH, $\rho
_{gas}$ is the gas density, $G$ is the gravitational constant. The accretion
rate of the black hole is limited to the Eddington value. A portion,
represented by $\varepsilon \dot{M}_{accr}$, of the mass being accreted is
transformed into radiation. Therefore, the effective rate of growth of the
black hole's mass can be expressed as $M_{BH}=\left( 1-\varepsilon \right) 
\dot{M}_{accr}.$

In the following, we will apply the accretion laws in the context of PBHs
that we assume as seeds for the observed AGNs. Similarly to \cite{E35}, we
suggest that the bolometric luminosity, denoted as%
\begin{equation}
L_{bol}=\varepsilon \dot{M}_{PBH}c^{2},
\end{equation}%
is evenly distributed thermally and isotropically among the gas particles
within the smoothing volume surrounding the PBH. The Eddington ratio \cite%
{E36}, denoted as $\Gamma =L_{bol}/L_{Edd}$, represents the relationship
between the bolometric luminosity of a quasar and its Eddington luminosity.
This ratio is formulated under the assumption of hydrostatic equilibrium and
pure ionized hydrogen and can be expressed as:%
\begin{equation}
L_{Edd}=\frac{4\pi M_{PBH}m_{p}c}{\sigma _{T}},
\end{equation}%
here $m_{p}$ is the mass of a proton and $\sigma _{T}$ is the Thomson
scattering cross-section. In this research, we proposed that the BHL
accretion model offers a potential explanation, wherein PBHs accrete
surrounding gas from the interstellar medium at a rate dependent on their
mass, relative velocity, and local gas density. This model, combined with
the Eddington limit which sets an upper bound on the luminosity and thus the
accretion rate based on the balance between radiation pressure and
gravitational pull can describe the rapid growth phases of these black
holes. If PBHs can accrete at near-Eddington or even super-Eddington rates
under certain conditions, such as dense gas environments or during periods
of high gas inflow, it would help explain how these distant SMBHs
accumulated mass so quickly. By applying the Eddington accretion rate and
the bolometric luminosity which are directly proportional to the efficiency
factor $\varepsilon $ and PBH mass, we can better understand the dynamics
and efficiency of early black hole growth, potentially resolving the puzzle
of their rapid formation and contributing to our knowledge of the early
universe's evolution.

\begin{figure*}[h]
\resizebox{1\textwidth}{!}{  \includegraphics{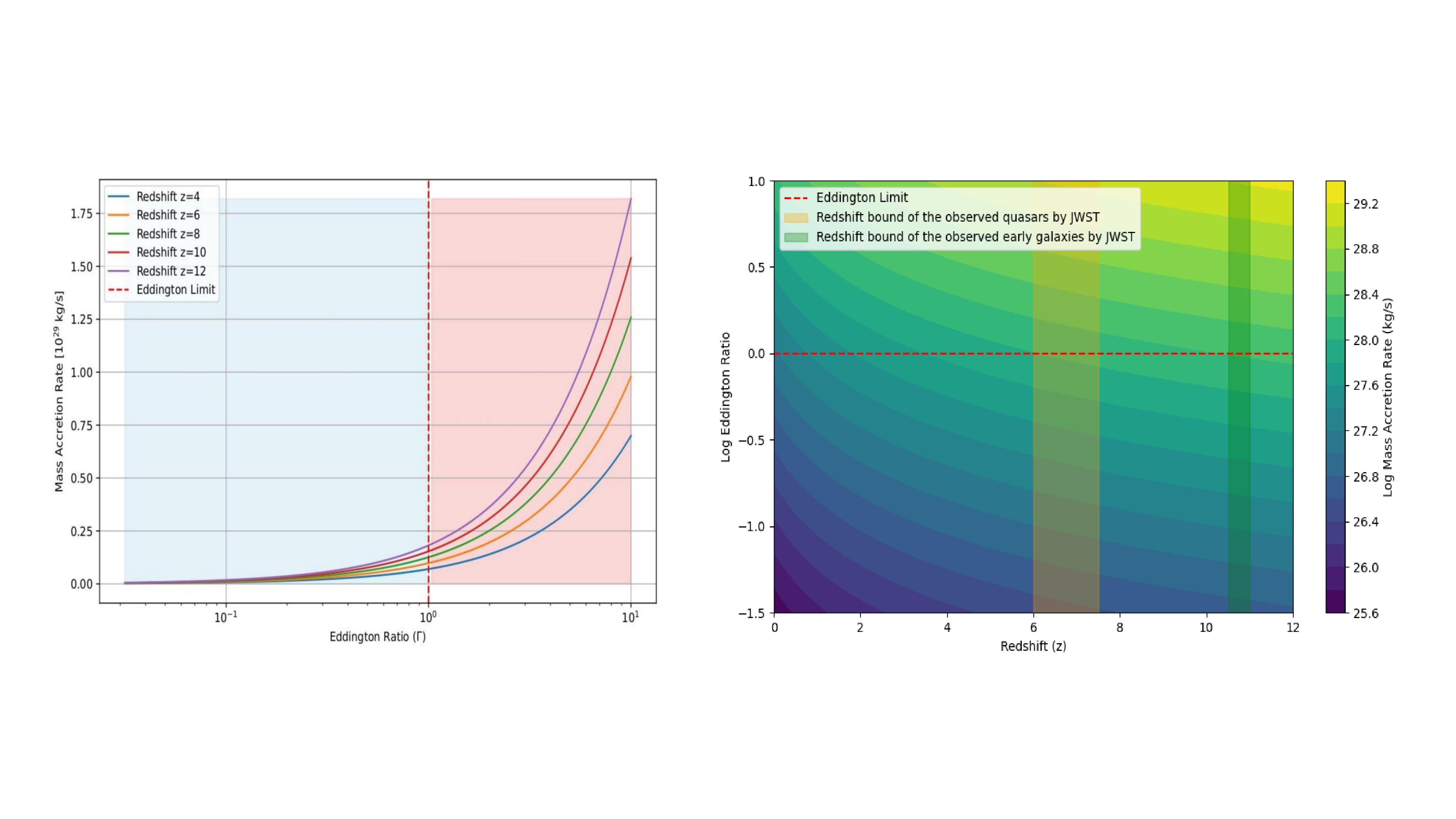}}
\caption{ The mass accretion rate of primordial black holes as a function of
the Eddington ratio for different redshifts.}
\label{fig:8}
\end{figure*}

Fig. \ref{fig:8} shows the mass accretion rate onto PBHs as a function of
the Eddington ratio $\Gamma $ across different redshifts $z$, reflecting
various epochs in the early universe. The results are plotted on a
logarithmic scale. The left side of the plot shows the relationship between
the Eddington ratio and the accretion rate, with shaded regions, the red one
highlights values below the Eddington limit and the blue region represents
values above the Eddington limit, the results are plotted for different
values of the redshift $z\sim \left[ 4,12\right] ,$ we noticed that the
accretion rate of PBHs as a function of the ratio $\Gamma $ is only
noticeable around $\dot{M}_{PBH}\propto 10^{29}kg/s.$\ Moreover, increasing
the redshift values makes the accretion rate faster when crossing the
Eddington limit $\Gamma \sim 1.$\ On the right side, The accretion rate is
computed for a range of Eddington ratios and redshifts, forming a
two-dimensional grid of values. The results are displayed in a contour plot,
showing the logarithm of the mass accretion rate across different redshifts
and Eddington ratios. An Eddington limit line is included to indicate the
boundary between sub-Eddington and super-Eddington accretion. Additionally,
colored regions highlight specific redshift bounds where early galaxies and
quasars have been observed by the JWST. Following these results we conclude
that the accretion of PBHs can be considered as sub-Eddington when we
consider values below $\dot{M}_{PBH}\lesssim 10^{28}kg/s$ for the lower
bound on the redshift that corresponds to quasars observations and the
higher bound related to early Galaxies observations, while both redshift
regions are considered super-Eddington for $\dot{M}_{PBH}\gtrsim 10^{28}kg/s.
$ \ 
\begin{figure*}[h]
\resizebox{0.85\textwidth}{!}{  \includegraphics{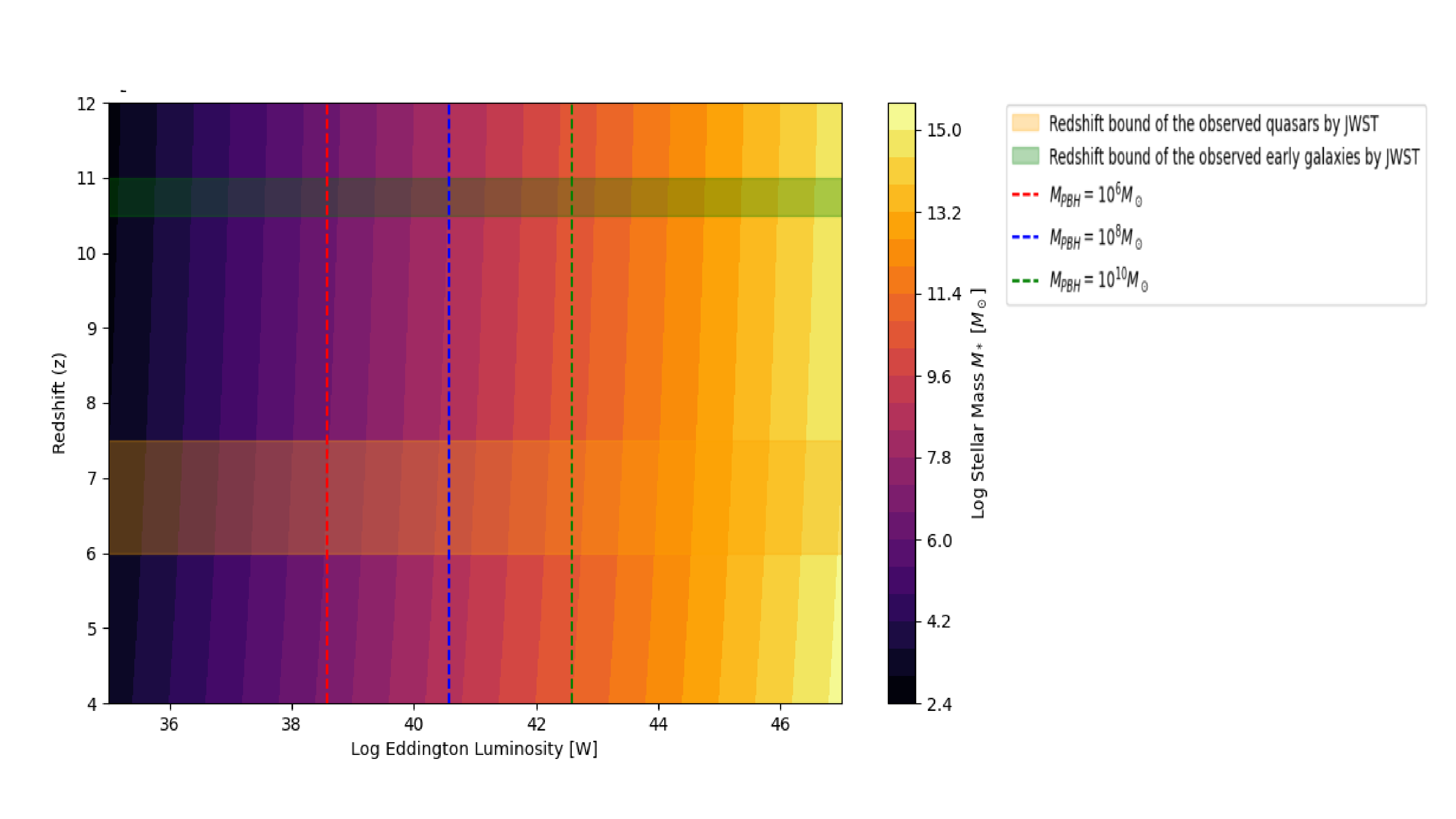}}
\caption{ Stellar Mass as a Function of Logarithmic Eddington Luminosity and
Redshift Across Different Cosmic Epochs.}
\label{fig:9}
\end{figure*}

Fig. \ref{fig:9} visualizes galaxies' stellar masses required to achieve
certain Eddington luminosities across various redshifts. The results are
presented in a contour plot, showing the logarithm of the stellar mass as a
function of logarithmic Eddington luminosity and redshift, with additional
vertical lines indicating specific limits of Eddington luminosity values of
PBHs of different masses. Shaded regions highlight redshift bounds
corresponding to observations of quasars and early galaxies by the JWST. For
both cases of the redshift bounds, we found that the Eddington limit can
increase when taking higher values of PBHs masses. In fact, this model can
provide a much clearer idea on the sub- or super-Eddington luminosity of
stellar masses taking into account the redshift values and PBHs masses, we
conclude that for the lowest chosen PBH masse $M_{PBH}\propto 10^{6}M_{\odot
}$ the Eddington limit can reach $L_{Edd}\sim 10^{38.5}W\sim 10^{11}L_{\odot
},$ while for the highest chosen PBH masse $M_{PBH}\propto 10^{10}M_{\odot }$%
\ the Eddington limit became $L_{Edd}\sim 10^{42.5}W\sim 10^{15}L_{\odot }.$%
\ This visualization helps in understanding the conditions under which PBHs
could grow rapidly, potentially explaining the formation of the most distant
and massive black holes observed, and providing insights into the accretion
dynamics and the role of PBHs in cosmic evolution. 
\begin{figure*}[h]
\resizebox{1.08\textwidth}{!}{  \includegraphics{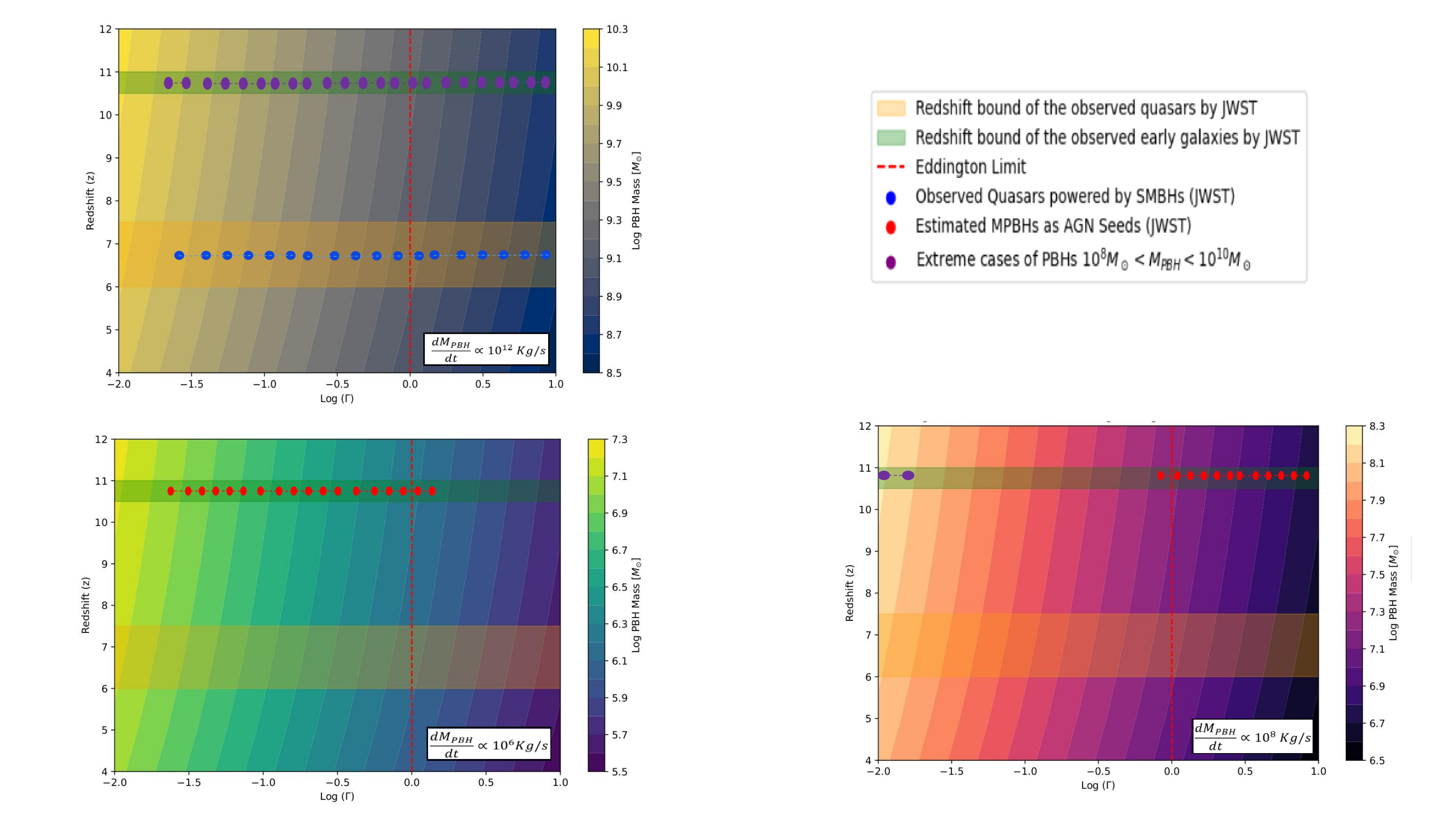}}
\caption{ Primordial Black Hole Mass Across Eddington Ratios and Redshifts.}
\label{fig:10}
\end{figure*}

Fig. \ref{fig:10} investigates the relationship between the mass of PBHs,
their Eddington ratios, and redshifts. The results are displayed in a
contour plot, showcasing the logarithm of PBH masses across different
Eddington ratios and redshifts. This visualization offers insights into the
growth potential of PBHs in the early universe and how their masses vary
with changing accretion rates and cosmic epochs. Additionally, shaded
regions highlight redshift bounds corresponding to observations of early
galaxies and quasars by the JWST, emphasizing the relevance of the study to
current observational constraints. The Eddington limit line shows the
boundary between sub-Eddington and super-Eddington regimes, providing
context for the efficiency of PBH accretion. Furthermore, the blue dots
represent the observed quasars powered by SMBHs based on recent
observations. It also includes red dots which symbolize observations of
GN-z11 AGN, suggesting these might be powered by MPBHs. Finally, the purple
dots denote extreme cases of PBHs with masses between $10^{8}M_{\odot }$ and 
$10^{10}M_{\odot }$. From these results we conclude that the choice of the
accretion rate plays an essential role in understanding galaxy formation and
black hole growth from different eras in our Universe, we can also see that
lower PBHs accretion speed of around $\propto 10^{6}kg/s$\ can enable the
existence of the estimated MBHs of AGNs like GN-z11 on sub-Eddington rate
for masses $M_{PBH}\lesssim 10^{6}M_{\odot },$ and super-Eddington for
higher masses. Moreover, the higher the accretion rate became the earlier
MBHs and SMBHs can exist either in the early Universe or subsequent eras. In
fact, an accretion speed around $\propto 10^{8}kg/s$\ can make the process
of MBH's AGNs on sub-Eddington regime, and SMBHs formation possible with
super-Eddington rate for masses $M_{PBH}\gtrsim 10^{7}M_{\odot },$ while for
accretion speed $\dot{M}_{PBH}\propto 10^{12}kg/s,$ it is possible to have
SMBHs that powers the observed quasars around $z\sim 6-7$ with sub-Eddington
rate for masses $M_{PBH}\lesssim 10^{9}M_{\odot }$\ and super-Eddington rate
for higher masses, while extream cases of PBHs masses can co-exist on the
same rate of accretion condition and masse constraints. Overall, this
program contributes to understanding the formation and evolution of PBHs in
the early universe, offering valuable insights into their role in cosmic
structure formation and the growth of supermassive black holes.

\section{Conclusion}

\label{sec6}

The process of B-L symmetry breaking at the Grand Unified Theory scale can
be counted as an additional factor for the initial conditions of the
Universe, which affects the entropy, the occurrence of baryogenesis through
leptogenesis, and the formation of dark matter due to the thermal generation
of gravitinos. In the preceding sections, we computed the gravitational wave
pattern resulting from the cosmological breaking of B-L symmetry. This
process occurred due to several factors, including the era of inflation, the
phase transition of the Universe, Cosmic strings, and the gravitational
waves linked to binary systems of black holes forming in the early Universe.
Additionally, our focus was directed toward the JWST predictions related to
the PBH masses and the early galaxy formation.

To go into further detail, our findings show first that considering the
scalar potential governing the B-L symmetry breaking mechanism, the model
parameter $\lambda $\ shows good consistency for a wide range of model
parameters with the latest observations from Planck data. As for the
critical value of the phase transition parameter $\upsilon _{B-L},$\ the
results show that it must be bounded around $\sim [2\times 10^{15}, 5\times
10^{15}] GeV,$ our findings also align with earlier findings that suggested
tuning the $\upsilon _{B-L}$ value to be close to the GUT scale. Moreover,
when we study the evolution of the B-L symmetry breaking parameter $\upsilon
_{B-L}$ as a function of the wavelength $k,$ we conclude that considering
superhorizon scales the potential parameter will decreased as we take higher
values of inflation e-fold number. Additionally, the spectral index shows
that for both smaller and larger scales the $\upsilon _{B-L}$ is best
considered around the previously observed values. On the other hand, for the
variation of the density of primordial gravitational waves with respect to
the scale $k$ and\ $\upsilon _{B-L}$\ for different phases, we found that
the change in the critical value $\upsilon _{B-L}$ that corresponds to B-L
asymmetry has a proportional effects on $\Omega _{gw,0}(f)$ for all matter,
radiation, and kinetic eras, from the results, we see that the proposed
scenario of Leptogenesis has great implications of the observed values of
PGWs density which can be predicted by PTA signals.

Moreover, we studied the NanoGrav proposed model of the density $h^{2}\Omega
_{gw}(f)$ as a function of the frequency $f$, scalar-to-tensor ratio $r$,
and the amplitude of the PTA signal that relates to the proposed potential
parameters that explain the Leptogenesis phenomena, we identified the right
parameters that can predict higher or lower density of gravitational waves
that reproduced the NANOGrav predictions, the results show good consistency
with the predicted bounds on the density and the frequency once we fine-tune
the inflationary parameter with parameters associated with the PTA signals.

We have calculated the density of gravitational waves from BBH mergers that
can be predicted by PTA interpretation, we conclude that the BBH system can
partially contribute to the explanation of the energy spectrum at nanohertz
frequencies. Moreover, PBH mass around $M_{PBH}\propto 10^{9}M_{\odot }$
presents the best fit with recent NANOGrav data. Additionally, PBH binaries
are important stochastic gravitational wave background sources that are used
to probe the Universe and can provide a possible explanation for the recent
NANOGrav results and JWST predictions. On another hand, when we study PBH
formation $\upsilon _{B-L}$ has a notable effect on the behavior of the
fraction of the total energy density where they converge towards values $%
M_{PBH}\propto 10^{16}g.$\ However, the mass of PBH at the formation time
fail to explain recent data, for this reason, we need to consider that the
evolution of PBHs has already happened which leads to massive galaxies
formation, then supplementary criteria will be considered to explain the
observations made by the JWST. The variation of galaxies stellar masses $%
M_{\ast }$ as functions of the redshift $z$, has shown that the galaxy
formation efficiency has a notable effect on the PBH mass, then the
influence of the seed, effect remains a viable tool for interpreting the
JWST observations.

In our study, we apply the Bondi-Hoyle-Lyttleton (BHL) accretion model to
primordial black holes (PBHs) as seeds for the observed AGNs, suggesting
that PBHs can accrete surrounding gas at rates dependent on their mass,
relative velocity, and local gas density. This model, combined with the
Eddington limit which balances radiation pressure against gravitational pull
can describe the rapid growth phases of these black holes. Our results
indicate that PBHs can accrete at near-Eddington or super-Eddington rates
under specific conditions, such as dense gas environments or periods of high
gas inflow, potentially explaining the rapid mass accumulation of these
distant SMBHs. We conclude that the BHL accretion model, when integrated
with the Eddington accretion rate and bolometric luminosity proportional to
the efficiency factor $\varepsilon $ and PBH mass, provides insights into
the dynamics and efficiency of early black hole growth. This could help
resolve the puzzle of their rapid formation and enhance our understanding of
the early Universe's evolution.

\section*{Acknowledgements}

The work by M.K. was performed with the financial support provided by the Russian Ministry of Science and Higher Education, project “Fundamental and applied research of cosmic rays”, No.~FSWU-2023-0068..

\end{document}